# Graphene made easy: high quality, large-area samples

Abhay Shukla*, Rakesh Kumar, Javed Mazher and Adrian Balan

*Université Pierre et Marie Curie-Paris6, CNRS-UMR7590, IMPMC, 140 rue de Lourmel, Paris, F-75015 France*

**We show that by using an original method, bulk graphite can be bonded onto borosilicate glass or potentially any insulating substrate with ionic conductivity and then cleaved off to leave single or few layer graphene on the substrate, identified optically and with Raman spectroscopy. This simple, inexpensive and fast method leads to the preparation of large area graphene and single or few-layer films of layered materials in general. We have prepared mm size few-layer graphene samples and also measured I-V characteristics in a FET. This opens up perspectives both for fundamental research as well as for applications.**

The demonstration[1-2] that graphene, a single atomic layer of graphite, can be obtained and manipulated has provoked a tremendous surge of work in this field, concerning both the singular properties of graphene as well as their applications in devices[3]. Though other methods have been recently reported[4], two methods have been used with considerable success for obtaining graphene: the first uses annealing of a SiC substrate to create a few layer graphene surface[1] while the second uses micro-mechanical cleaving[2]. However there are limiting aspects to each of these methods. In the case of micro-mechanical cleaving these include sample size and inadaptability to large scale applications. In the case of annealed SiC, there are questions regarding sample quality[5] and the method is relevant only to graphene. We have developed a simple, inexpensive and fast method for producing larger graphene samples, and two dimensional samples of layered materials in general. This method is inspired from a technique used for bonding Si to a Pyrex substrate, known as anodic bonding [6,7]. By applying a potential difference of the order of a kV to a heated Pyrex/Si interface very intimate contact is obtained between the substrate and silicon which translates into the formation of chemical bonds at the interface. In the case of the Si/Pyrex interface the bond is permanent and irreversible. The principle of anodic bonding in Si/Pyrex is the following: At high temperatures (200°C-400°C), $Na_2O$ in Pyrex glass decomposes into $O^{2-}$ and $Na^+$ ions. The smaller $Na^+$ ions, being relatively mobile at these temperatures, migrate across the glass substrate under the influence of the applied electric field[8] towards the cathode at its back. The static oxygen ions left behind create a negative space charge on the glass surface at the interface and a high electrostatic field with the positively charged Si wafer. This field leads to intimate atomic contact between the substrate and the wafer and ultimately the formation of Si-O-Si bonds. The surface quality of the wafer and the substrate is important to establish a good bond.

It is known that conducting materials which readily oxidize can be bonded to Pyrex substrates through a similar mechanism, though the quality of the bond is variable. In the case of graphite it is probable that the initial bonding is predominantly electrostatic in nature since Raman spectra do not show any change that could be traced to the formation of graphene oxide, nor do they show any significant creation of defects. Needless to say, the temperature and the high voltage with the correct bias are essential and without them we have no adhesion and consequently no graphene on our substrate. Figures 1 and 2 show optical images of some graphene samples obtained by our method. In Fig. 1(a) we show the footprint left by a graphite flake, bonded onto a Pyrex substrate and then cleaved off. The brightest areas of the bonded film correspond to graphitic material. The fainter region to the left corresponds to a large trilayer of graphene with dimensions of more than 1mm x 0.5mm. This trilayer graphene film, visible to the naked eye, also contains two bilayer 'islands', one of which is shown in the blow-up in Fig. 1(b). The bilayer portions are strip like with dimensions of 300 μm x 50 μm and 250 μm x 70 μm. In Fig. 2(a) we show a monolayer sample with dimensions of about 200 μm x 150 μm containing holes, the largest continuous portion being about 150 μm x 50 μm. In Fig. 2(b) we show another monolayer sample with dimensions of about 120 μm x 70 μm. These are identified by means of the characteristic Raman signal but also using AFM imaging. The size and quality of samples depends both on the tuning of fabrication parameters and the choice of the precursor material. Natural graphite precursor results in bigger graphene than when HOPG graphite precursor is used.

In our setup we use from 1.2 to 1.7 kV with the anode on the graphite sample and the cathode contacting the backside of the pyrex substrate. The glass substrate with the graphite mounted on it is heated so that at the interface the temperature is around 200°C. With the temperature stabilized, the potential difference is applied and a current is immediately detected signalling the migration of the sodium ions inside the glass substrate. The peak current which is attained within a few seconds depends on the interface size and properties (cleanliness, flatness) and may vary from a few tens of μA to a few hundred μA. As the space charge layer is formed and begins to counteract the applied voltage, the current decreases exponentially. We generally achieve bonding within a few minutes both with 500 μm thick Pyrex-7740 substrates as well as ordinary 120 μm thick laboratory borosilicate cover-slips. After bonding is achieved the bulk graphite sample can be cleaved off, leaving several bonded areas on the glass surface. These are then peeled off using adhesive tape to leave many transparent areas with one-layer, or few-layer graphene portions. It should be noted that all these steps are performed in standard laboratory environment, there being no need at this stage for a controlled, clean room atmosphere. We pre-identify the graphene layers with optical microscopy. In Fig. 1 and Fig. 2 few layer and monolayer graphene portions are visible as faintly contrasted areas with respect to the substrate, the contrast increasing with the number of graphene layers. This possibility of visual identification, earlier established in the case of $SiO_2$ on Si substrates[2], makes for straightforward manipulation of these samples and we can then proceed to rigorous identification of the number of layers through micro-Raman spectroscopy. We have measured the characteristic peaks for graphene with 514 nm wavelength incident photons and a spot size of about 1 μm. In Fig. 3 we show Raman spectra from four random spots taken over the large trilayer region, from the bilayer samples and from the monolayer samples shown in Fig. 1 and Fig. 2. The Raman spectra show the 2D (~2700cm$^{-1}$) peak, corresponding to a double resonant scattering process from zone-edge phonons and the G peak (~1590cm$^{-1}$) which is a doubly degenerate zone centre $E_{2g}$ mode. The microscopic process responsible for the 2D peak is such that the line shape depends only on the band structure of the few layer graphene[9]. The 2D peak lineshape is thus a very reliable marker for distinguishing between different numbers of layers in graphene samples as opposed to the relative G/2D intensity which depends on a number of parameters which ultimately influence the effective doping level[10]. We remark that the D peak (~1350cm$^{-1}$) which indicates the presence of defects is entirely absent or very weak in our samples, testifying to their excellent quality. We have also mapped some regions of our samples using atomic force microscopy and thus confirmed the information obtained from Raman spectroscopy pertaining to the number of layers. In particular, for the sample shown in Fig. 1, we find a step-size of 1.5-1.8 nm between trilayer and substrate, a step-size of 1.1- 1.3 nm between bi-layer and substrate and a step-size of 0.5 nm between bilayer and trilayer. When one considers that the substrate roughness is about 0.4 nm (RMS) and that graphene to substrate step-sizes measured by AFM are in general more than what one would expect while layer step-sizes on graphite are accurate[2], these values corroborate very well with our Raman measurements.

Finally in Fig. 4 we show device characteristics of a top-gated bilayer graphene sample made by our method. The sample was patterned using photolithography and Au/Cr (30/10 nm) bilayer deposition. We measure a resistance of about 1 k$\Omega$ between nearest neighbor contacts. By measuring the resistance as a function of distance between contacts the contact resistance is estimated to be below 100 $\Omega$. These values compare very favorably with values reported for graphene made with micro-mechanical cleaving[10,11]. We have undertaken electrochemical top-gating with a solid (Polyethylene oxide/LiClO$_4$) electrolyte[10] which allows high doping levels with low gate voltages. On the addition of the electrolyte the sample resistance increases, possibly due to the addition of impurities as reported in reference 10. In Fig. 4 we show the square resistance of our bilayer FET (calculated using the aspect ratio width/length =3.2 for the pair of contacts used) and in the inset, the extracted mobility as a function of the doping concentration estimated from the electrochemical gate voltage[10] for the two point probe measurement configuration. The charge neutrality peak, characteristic of few-layer graphene, is found at a gate voltage of -0.3 V, so at practically zero doping, implying that we have no parasite doping from the anodic bonding process. I-V characteristics between source and drain show linear behaviour excluding Schottky barriers at the contacts, and expectedly reflect the drop in resistance due to the field effect doping.

In conclusion, promising perspectives are opened up by our method. Improvements in our process and eventually in the quality of the bulk graphite precursor should lead to bigger graphene samples for use in devices. Our process could be applied to other, technologically relevant substrates where relatively mobile ions can be produced under certain conditions. A thin layer of borosilicate glass on Si, with an eye on device making, is a distinct possibility. Lastly this method can be used to make few-layer and monolayer samples of other layered materials and we have succeeded in manufacturing few layer samples of $NbSe_2$ and InSe. In particular semi-conducting monolayers could be immediately relevant for micro-electronic applications.

**Acknowledgements**

We acknowledge J-C Bouillard, M. D'Astuto and F. Mauri for discussions, F. Gelebart, M. Morand, P. Munsch, J-C Chervin, E. Lacaze and R. Gohier for help. J.M. was financed by the Indo-French Center for the Promotion of Advanced Research. The process described in this work has been deposited as INP patent no. 0707145.

\* Correspondance and requests for materials should be addressed to A.S. (abhay.shukla@impmc.jussieu.fr).

**Figure Captions**

FIG. 1. (Color online) Optical micrographs with polarized reflected light showing regions of few layer graphene bonded onto the glass substrate. The brighter portions correspond to thicker layers, while the weakest contrast is from the few layer graphene portions. Shown inside the areas demarcated by rectangles: (a) a big mm sized trilayer film (b) a bilayer island in the upper left portion of the trilayer film. The visible, 150 μm spaced grid was deposited for ulterior photolithography.

FIG. 2. (Color online) Optical micrographs with polarized reflected light showing regions of monolayer graphene (weakest contrast) bonded onto the glass substrate. (a) a large, isolated but discontinuous sample (b) a smaller continuous sample between thicker regions, with some sharp edges.

FIG. 3. (Color online) Raman spectra of samples shown in Figs. 1 and 2. The G peak (left panels) at ~1590 cm$^{-1}$ and the 2D peak (right panels) at ~2700 cm$^{-1}$ in graphene taken with 514 nm wavelength incident light are shown. The D peak intensity (at ~1350 cm$^{-1}$) corresponds to the defect level and by its near absence testifies to the excellent quality of our samples. The G peak is a zone centre, in-plane phonon. The 2D lineshape, arising from a double resonant scattering process from a zone edge phonon is a reliable indication of the number of layers. Top: two monolayer graphene samples shown in Fig 2a and 2b; Middle: two bilayer graphene samples including the one shown in Fig 1b; Bottom: Four random points on the trilayer graphene sample of Fig 1a.

FIG. 4. (Color online) Resistivity measurements in a two point, top-gated configuration. Top panel: The horizontal contacts of the sample were used (left inset, sample shown

before top-gating, the distance between these contacts is 4 μm). The variation of the square resistance is shown as a function of the top gate voltage and the right inset shows the extracted mobility as a function of doping. Bottom panel: I-V characteristics as a function of the gate voltage.

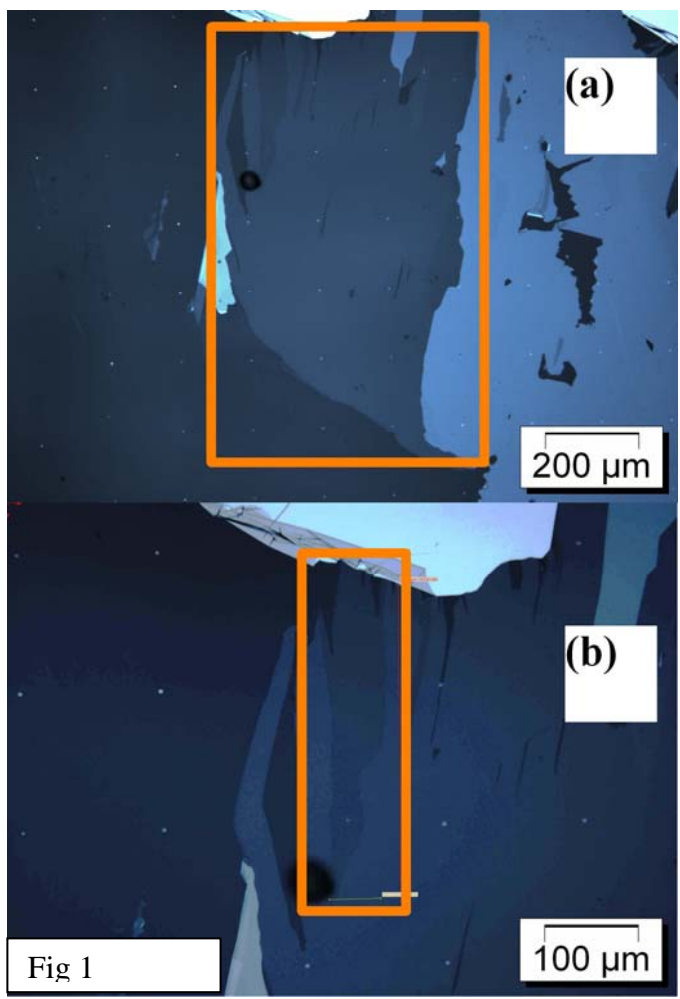

**(a)**

200 μm

**(b)**

100 μm

Fig 1

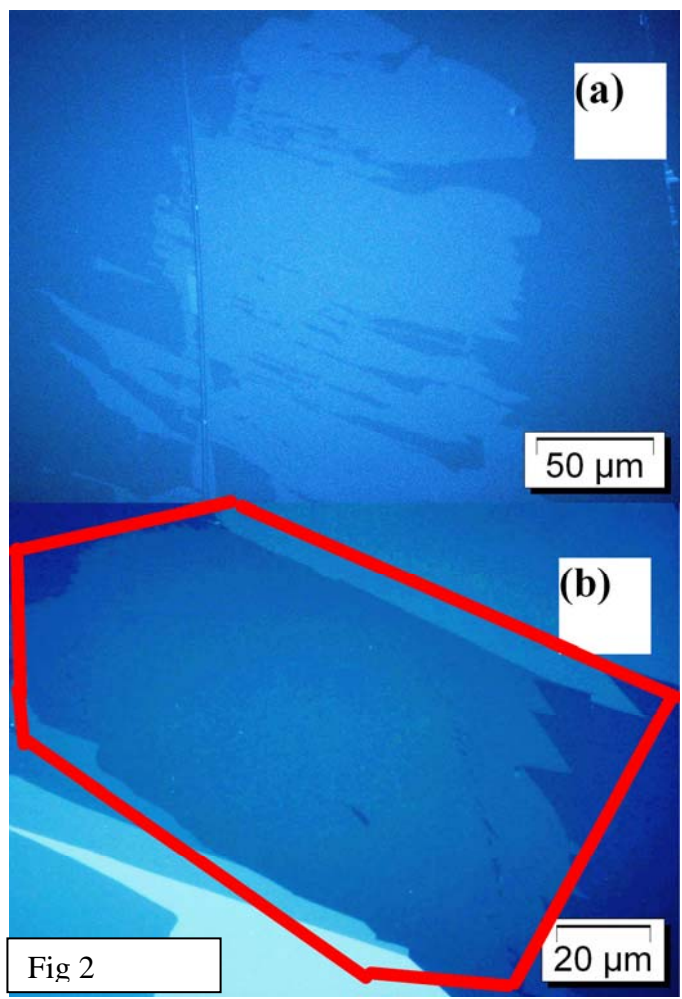

**(a)**

50 μm

**(b)**

20 μm

Fig 2

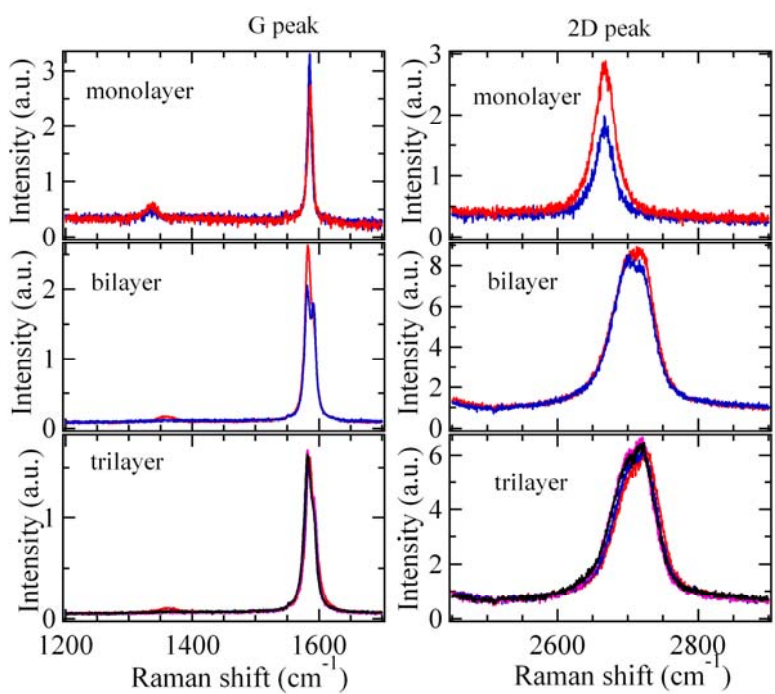

Fig 3

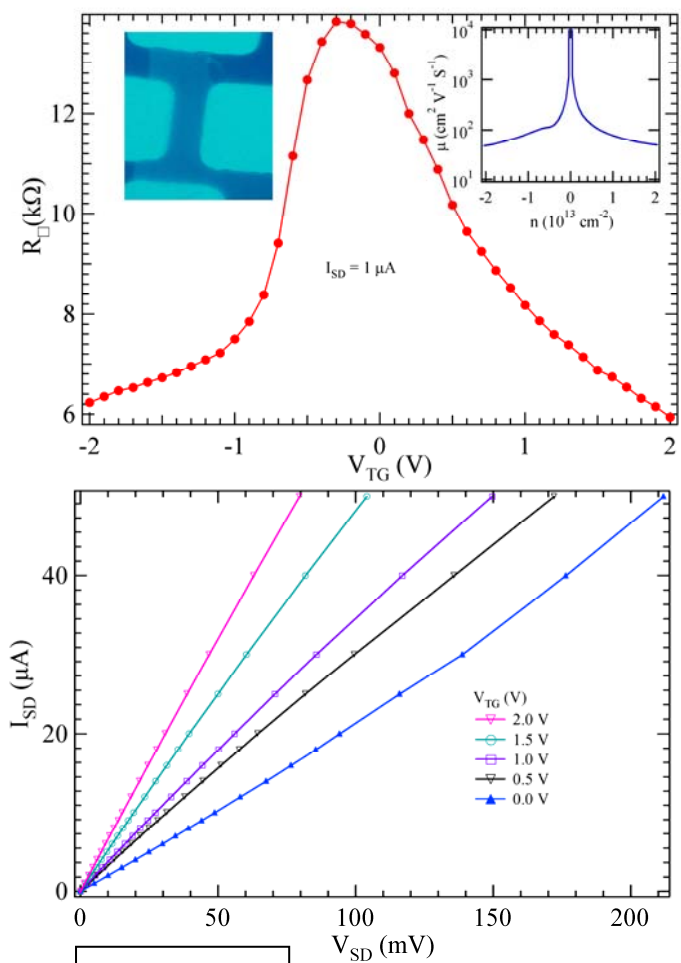

Fig 4